\documentclass[reprint,a4paper,aps,prl,amsmath,amsfonts,showpacs,floatfix,nofootinbib]{revtex4-1}
\usepackage{bm}                      
\usepackage{graphicx}                
\usepackage{dcolumn}                 
\usepackage[table]{xcolor}
\usepackage{amsmath}
\usepackage{times}

\allowdisplaybreaks

\newcommand*{\centt}[1]{\multicolumn{1}{c}{#1}}
\newcommand*{\cent}[1]{\multicolumn{1}{c}{$#1$}}
\newcolumntype{x}[1]{D{.}{.}{#1}}

\newcommand{\icm}{\text{cm}^{-1}}

\begin{document}
\preprint{Version 2.0}

\title{Nonadiabatic QED Correction to the Dissociation Energy of the Hydrogen Molecule}

\author{Mariusz Puchalski}
\affiliation{Faculty of Chemistry, Adam Mickiewicz University, Umultowska 89b, 61-614 Pozna{\'n}, Poland}

\author{Jacek Komasa}
\affiliation{Faculty of Chemistry, Adam Mickiewicz University, Umultowska 89b, 61-614 Pozna{\'n}, Poland}

\author{Pawe\l\ Czachorowski}
\affiliation{Faculty of Physics, University of Warsaw, Pasteura 5, 02-093 Warsaw, Poland}

\author{Krzysztof Pachucki}
\affiliation{Faculty of Physics, University of Warsaw, Pasteura 5, 02-093 Warsaw, Poland}

\date{\today}

\begin{abstract}

  The quantum electrodynamic correction to the energy of the hydrogen molecule
  has been evaluated without expansion in the electron-proton mass ratio.
  The obtained results significantly improve the accuracy of theoretical predictions
  reaching the level of 1 MHz for the dissociation energy, in a very good agreement
  with the parallel measurement [H\"olsch {\em et al.}, Phys. Rev. Lett. {\bf 122}, 103002 (2019)].
  Molecular hydrogen has thus become a cornerstone of ultraprecise quantum chemistry,
  which opens perspectives for determination of fundamental physical constants from its spectra.

\end{abstract}
\maketitle

{\sl Introduction} --
The spectra of hydrogenic atoms are being used for determination of physical constants
and for precision tests of the fundamental interactions theory.
However, the finite lifetime of excited states makes further progress in accuracy very challenging.
It would require, among others,
determination of the resonance frequency to at least one part
in 10,000 of the observed line width \cite{Beyer:17}.  
In contrast, the hydrogen molecule (H$_2$) has many narrow lines, which in principle
can be measured very accurately \cite{Liu:09,Sprecher:11,Cheng:12,Niu:14,Cheng:18,Altmann:18}.
In this work, we demonstrate that they can also be calculated very accurately,
namely with 1 MHz uncertainty or better.

Although the hydrogen molecule is one of the simplest molecular systems,
the high-precision calculations of its energy levels have been difficult to perform, 
even in the nonrelativistic limit. The standard Born-Oppenheimer (BO) approximation gives 
a relative accuracy of the order of $10^{-3}-10^{-4}$ only as a consequence of the omission 
of the coupling between electrons and nuclei movements.
In principle, the finite nuclear mass corrections to the BO potential can be included
systematically within the nonadiabatic perturbation theory (NAPT) \cite{PK09}.
However, evaluation of the higher order terms of the NAPT becomes complicated \cite{PK15}. 
For this reason the direct nonadiabatic methods have recently been developed in which
two electrons and two protons are treated on the same footing.
This allowed the inaccuracy of the nonrelativistic energy $E^{(2)}$ 
to be reduced to the limit of $4\times 10^{-12}$ resulting from the uncertainty in the proton mass  \cite{PK:16}.

Regarding subsequent terms in the expansion of energy in the fine structure constant 
$\alpha$, 
\begin{equation}\label{alphaexp}
  E(\alpha) = \alpha^2\,E^{(2)} +  \alpha^4\,E^{(4)} +  \alpha^5\,E^{(5)} +  \alpha^6\,E^{(6)} +
              \alpha^7\,E^{(7)} + \ldots ,
\end{equation}
the relativistic correction, $E^{(4)}$, has recently been calculated to a high numerical precision 
both with direct nonadiabatic treatment \cite{Wang:18a,PSKP:18,Wang:18b} 
and with the NAPT \cite{PKP:17,CPKP:18}. 
Moreover, the quantum electrodynamic (QED) $E^{(5)}$ \cite{PLPKPJ09} 
and the higher order $E^{(6)}$ \cite{PKCP:16} corrections
had been calculated within the BO approximation only,  while $E^{(7)}$
is known approximately from the atomic hydrogen theory \cite{Eides:01}. 
Neglected nonadiabatic effects of the order $O(\alpha^5)$ had been the largest source 
of the uncertainty in theoretical predictions, which
for the ground state dissociation energy equals to $2\times 10^{-4}\,\icm$ (6 MHz) \cite{PSKP:18}. 
 
In this work, we report on calculation of the QED correction $E^{(5)}$ 
in the framework of the direct nonadiabatic approach,
improving the accuracy of theoretical prediction for the dissociation energies
by an order of magnitude down to $2.6\times 10^{-5}$ cm$^{-1}$ (0.78 MHz), the best ever
theoretical prediction for any molecule, which becomes now sensitive to the nuclear charge radii.

{\sl Nonrelativistic wave function} --
The basis of the accurate theoretical predictions is the precise nonrelativistic wave function.
In the nonadiabatic approach all particles are treated on an equal footing, and the wave function
is an eigenstate of the nonrelativistic H$_2$ Hamiltonian of the form
\begin{equation} \label{Ham}
H = T + V\,,
\end{equation}
where
\begin{align}
T &=\frac{ \vec p_0^{\,2}}{2\,m_p} + \frac{ \vec p_1^{\,2}}{2\,m_p} +  \frac{\vec p_2^{\,2}}{2 \, m} + \frac{\vec p_3^{\,2}}{2\, m}\,,  \\
V &=\frac{1}{r_{01}}-\frac{1}{r_{02}}-\frac{1}{r_{03}}-\frac{1}{r_{12}}-\frac{1}{r_{13}}+\frac{1}{r_{23}}\,. 
\end{align}
The indices 0,1 denote protons of mass $m_p$ and 2,3---electrons of mass $m$.
In the center of mass frame the nonrelativistic wave function depends only on the interparticle distances $r_{ij}$.
The most convenient approach \cite{PSKP:18} to calculate this function
is based on a variational principle with the nonadiabatic explicitly correlated Gaussian (naECG) functions.
In this approach, the wave function is represented as
\begin{eqnarray}
 \Psi &=& \sum_i^N c_i\, \psi_i(\vec r_0, \vec r_1, \vec r_2, \vec r_3)\,, \\
 \psi_i &=& (1 + P_{0\leftrightarrow 1})\,(1+P_{2\leftrightarrow 3})\,\phi_i (\vec r_0, \vec r_1, \vec r_2, \vec r_3), \label{psii}
 \end{eqnarray}
where $P_{i\leftrightarrow j}$ is the particle exchange operator, which accounts for the fact that the ground state H$_2$
wave function $\Psi$ is symmetric with respect to the exchange of nuclear and electronic variables.
The spatial functions $\phi_i$ in Eq.~(\ref{psii}) are naECG functions of the form
\begin{equation}
\phi =r_{01}^n\, e^{-a_1 r^2_{01}-a_2 r^2_{02}-a_3 r^2_{03}-a_4 r^2_{12}-a_5 r^2_{13}-a_6 r^2_{23}}\,. \label{phi} 
\end{equation}
The nonlinear $a$ parameters are optimized variationally and the internuclear coordinate 
$r_{01}^n$ prefactor ensures proper representation of the vibrational part of the wave function. 
The powers $n$ of this coordinate are restricted to even integers within the range $0-80$
and are generated following the log-normal distribution. 
The nonrelativistic wave function $\Psi$ has been optimized for several basis sizes
to observe the convergence of the nonrelativistic energy and expectation values.

{\sl QED correction} --
The formula for the leading quantum electrodynamic correction $E^{(5)}$, derived
here on the basis of QED theory in agreement with the known formulas for H \cite{Eides:01} 
and He \cite{Pachucki:00}, is
\begin{widetext}
\begin{align}
  E^{(5)} =&\ -\frac{2{\cal D}}{3\,\pi} \ln k_0
  -\frac{7}{6\,\pi} \biggl\langle \frac{1}{r_{23}^3}
  + \frac{m}{m_p}\sum_{a,x}\frac{1}{r_{ax}^3}
 + \frac{m^2}{m_p^2}\frac{1}{r_{01}^3} \biggr\rangle_{\!\epsilon}
 +\frac{4}{3}\,\biggl\{\frac{19}{30}+\ln(\alpha^{-2})
+\frac{m}{4\,m_p}\biggl[\frac{62}{3}+\ln(\alpha^{-2})\biggr]
\nonumber \\&\
+\frac{m^2}{m_p^2} \biggl[\ln\biggl(\frac{m_p}{m}\biggr)
  + \ln(\alpha^{-2}) + 4\biggr]
\biggr\}\sum_{a,x}\big\langle \delta^3(r_{ax}) \big\rangle
+\left(\frac{164}{15}+\frac{14}{3}\,\ln\alpha\right)\,\big\langle\delta^3(r_{23})\big\rangle - 2\,E^{(5)}_{\rm H}\label{E5mol}
\\
E^{(5)}_{\rm H} =&\ -\frac{4}{3\,\pi}\,\frac{\mu}{m}\,\biggl(\ln k_0^H+\ln\frac{\mu}{m}\biggr)
+\frac{1}{3\,\pi}\,\biggl(\frac{\mu}{m}\biggr)^3
\biggl\{4\,\biggl(1 + \frac{m}{4\,m_p} + \frac{m^2}{m_p^2}\biggr)\,\ln\bigl(\alpha^{-2}\bigr) + \frac{38}{15}
+ \frac{m}{m_p}\biggl(\frac{62}{3}+14\,\ln2\biggr)
\nonumber \\ &\ 
      + 2\,\frac{m^2}{m_p^2}\biggl(1+2\,\ln\frac{m_p}{m}\biggr)\biggr\},
\end{align}
\end{widetext}
where $a=0,1$, $x=2,3$, the Bethe logarithm is \cite{Drake:90}
\begin{align}
\ln k_0 =&\ \frac{1}{\cal D}\,
\left\langle \vec J \,(H-E)\,\ln\bigl[2\,(H-E)\bigr]\,\vec J \right\rangle\label{08} \\
\ln k_0^H  =&\ 2.984\,128\,555\,765\,498
\end{align}
and where
\begin{align}
\vec J &=\frac{\vec p_0}{m_p} + \frac{\vec p_1}{m_p} - \frac{\vec p_2}{m} - \frac{\vec p_3}{m}\,, \\
{\cal D}&=\left\langle \vec J \,(H-E)\,\vec J \right\rangle 
 = \frac{2\,\pi}{\mu^2}\,\sum_{a,x}\langle \delta^3(r_{ax}) \rangle\,, \\
\mu&= m_p\,m/\left(m_p + m \right).
\end{align}
In all the above formulas the expectation values are taken with the nonrelativistic wave function $\Psi$.
Moreover, the expectation values of singular terms $\langle 1/r^3\rangle_\epsilon$ are obtained 
by integration from $\epsilon$ to $\infty$ and subtraction of $\ln\epsilon + \gamma$,
where the symbol $\gamma$ indicates the Euler-Mascheroni constant. 

There are certain ambiguities regarding the molecular
QED correction of Eq.~(\ref{E5mol}), which need to be explained.
The first one is due to the lack of the contact term between protons $\delta^3(r_{01})$.
In fact, such a term exists, for example from the strong interaction $V_{\rm strong}$ between protons \cite{Wiringa:95}
\begin{align}
  \delta E_{\rm strong} =&\  \langle\delta^3(r_{01})\rangle\,\int d^3r V_{\rm strong}(r)
  \nonumber \\ =&\ \langle\delta^3(r_{01})\rangle\, (-2.389)\;{\rm fm}^2\,.
\end{align}
It is of the same order as the electron vacuum polarization correction to the
Coulomb interaction between protons
\begin{align}
  \delta E_{\rm vp} =&\,\langle\delta^3(r_{01})\rangle\,\Bigl(-\frac{4}{15}\Bigr)\,\frac{\alpha^2}{m^2}
  \nonumber \\ =&\ 
  \langle\delta^3(r_{01})\rangle\,(-2.118)\;{\rm fm^2}.
\end{align}
However, both contributions are totally negligible, because
$\langle\delta^3(r_{01})\rangle\sim 10^{-50}\,(m\,\alpha)^3$ for the ground state of H$_2$.

Another subtle point to be clarified is the proton self-energy correction 
and the corresponding definition of the proton charge radius. 
This correction is insignificant for a regular hydrogen atom but 
non-negligible for muonic hydrogen ($\mu$H). So, for consistency with
the determination of the proton charge radius $r_p$ in $\mu$H \cite{pohl:10}, 
we chose to include this effect into the total energy of H$_2$.
Following \citep{Pachucki:99}, we do so in a minimal way, by including only logarithmic terms,
and the nonlogarithmic terms are absorbed into the mean square charge radius $r_p^2$.

{\sl Numerical calculation} --
QED corrections involve several nontrivial terms:
the Bethe logarithm $\ln k_0$, the interparticle Dirac delta $\langle\delta(r_{ab})\rangle$, 
and the Araki-Sucher term $\langle 1/r_{ab}^{3}\rangle_{\epsilon}$. 
Because the naECG basis does not reproduce the electron-electron and electron-nucleus cusps 
of the wave function, the two latter terms exhibit a slow convergence if calculated directly 
from their definitions. Therefore, to increase numerical performance, it is crucial 
to transform singular operators to a more regular form, whose behavior is less sensitive
to the local inaccuracies of the wave function.  For this purpose, we generalized known relations
\cite{PCK:05,PKP:17} and obtained the following expectation value identities with $a,b=0,1,2,3$
\begin{align}
\label{deltareg}
\langle 4\,\pi\,\delta^3(r_{ab})\rangle  =&\ \frac{2\,m_a\,m_b}{m_a+m_b}
\\ &\ \nonumber \times
\bigg\langle\frac{2}{r_{ab}} (E-V)- \sum_c \frac{1}{m_c} \vec p_c\,\frac{1}{r_{ab}}\,\vec p_c \biggr\rangle,
\end{align}
\begin{align}
\label{rm3reg}
\bigg\langle\frac{1}{r_{ab}^{3}}\bigg\rangle_{\!\epsilon}  =&\
(1 + \gamma)\,\langle4\,\pi\,\delta^3(r_{ab})\rangle + \frac{2\,m_a\,m_b}{m_a+m_b}
\\ &\ \times
\bigg\langle\frac{2\,\ln r_{ab}}{r_{ab}} (E-V)
  - \sum_c \frac{1}{m_c} \vec p_c\,\frac{\ln r_{ab}}{r_{ab}}\,\vec p_c \biggr\rangle, \nonumber
\end{align}
\begin{align}
\bigg\langle\frac{1}{r_{ab}^4}\bigg\rangle_{\!\epsilon} =&\ \frac{m_a\,m_b}{m_a+m_b}\, 
\bigg\langle \sum_c \frac{1}{m_c} \vec p_c \frac{1}{r_{ab}^2} \vec p_c
\\&
- 2\,(E - V)\,\frac{1}{r_{ab}^2} \pm 12\,\pi\,\delta^3(r_{ab})\bigg\rangle_{\!\epsilon}. \nonumber
\end{align}
where $+$ is for particles with the same and $-$ with  opposite charges respectively. 
Results for the expectation values extrapolated to the infinite
basis size, along with their estimated uncertainty, are presented in Table~\ref{Toperators}.
\begin{table}[t!hb]
\renewcommand{\arraystretch}{1.2}
\caption{Expectation values of the operators present in Eq.~(\protect\ref{E5mol}).
Atomic units are used throughout the Table. }
\label{Toperators}
\begin{ruledtabular}
\begin{tabular}{l@{\extracolsep{\fill}}x{3.15}}
\centt{Operator} & \centt{Expectation value} \\
\hline\\[-1.5ex]               
$E$   
& -1.164\,025\,030\,86(3) \\
$\langle J^{\,2}\rangle$ & 2.518\,270\,507\,19(12) \\
$4\pi\sum_{a,x} \langle\delta^3(r_{ax})\rangle$
    & 11.346\,476\,34(9) \\
$4\,\pi\,\langle\delta^3(r_{23})\rangle $
    & 0.202\,830\,306(6) \\
$\sum_{a,x} \langle r_{ax}^{-3}\rangle_\epsilon$ 
    & -7.191\,104\,3(10) \\
$\langle r_{23}^{-3}\rangle_\epsilon$
    &  0.401\,943\,51(6) \\
$\langle r_{01}^{-3}\rangle_\epsilon$ 
    & 0.357\,215\,411\,7(3) \\
$\sum_{a,x} \langle r_{ax}^{-4}\rangle_\epsilon$ 
    &  -5.712\,727(17) \\
$\sum_{(a,x)<(b,y)} \Big\langle\frac{\vec r_{ax}}{r_{ax}^3}\cdot \frac{\vec r_{by}}{r_{by}^3}\Big\rangle $
    & -0.254\,515\,18(18)
\end{tabular}        
\end{ruledtabular}   
\end{table}

{\sl Bethe logarithm} --
Among all the terms in Eq. (\ref{E5mol}), the calculation of the Bethe logarithm $\ln k_0$
is the most complicated one. We express $\ln k_0$
in terms of the one-dimensional integral
\begin{equation}
\label{lnk0ft}
\ln k_0 = \frac{1}{\cal D}\,\int_0^1 dt\, \frac{f(t)-f_0 - f_2\,t^2}{t^3}  
\end{equation}
with the function $f(t)$ defined as
\begin{equation}
f(t) = \biggl\langle\vec J\,\frac{k}{k+H-E}\,\vec J\biggr\rangle, \qquad t = \frac{1}{\sqrt{1+2\,k}} 
\label{ft}
\end{equation}
which has the following Taylor expansion
\begin{equation}
f(t) = f_0 + f_2\; t^2 + f_3\; t^3 + (f_{4l} \ln t + f_{4})\; t^4  + O(t^5)\label{ftexp}\,.
\end{equation}
with the first two coefficients
\begin{equation}
f_0 = \langle J^2 \rangle, \qquad f_2 = -2\,{\cal D}\,.
\end{equation}
The integrand in Eq.~\eqref{lnk0ft}, as a smooth function of $t$ was evaluated at 200 equally
spaced points in the range $t \in [0, 1]$, which enabled relative accuracy higher than 
$10^{-7}$. In the numerical  calculation of $f(t)$, the resolvent in Eq. (\ref{ft}) was represented
in terms of pseudostates of the form  $\phi^{\Pi} =\vec{r}_{ab}\,\phi$ for all interparticle coordinates. 
The nonlinear parameters of $\phi^\Pi$ are found by a maximization of $f$.
In the particular case of $t=1$ $(k=0)$, $f$ can be evaluated analytically
using the generalized Thomas-Reiche-Kuhn sum rule \cite{Zhou:06}
$\langle \vec J \,(H-E)^{-1}\,\vec J \rangle= 3\,(1+m/m_p)$. We used this opportunity
to assess the completeness of the pseudostates space and to estimate uncertainties.

For the given size $N$ of the wave function $\Psi$ expansion, the size of pseudostate 
basis set was chosen as $N'=\frac{3}{2}N$, which appeared to be sufficient for most of the $t$ points. 
There are also additional factors taken into account for the accurate representation of
the resolvent in Eq. (\ref{ft}). 
The powers of internuclear coordinate $r_{01}$, analogously to the wave function, 
are restricted to even integers and are generated randomly for each basis function 
from the log-normal distribution within the $0-80$ range. 
However, for small values of $t$ ($\leq 0.1$), 
due to a cancellation in the numerator of Eq.~(\ref{lnk0ft}), 
an additional tuning of the distribution was made and $N'=2\,N$ was set
to achieve high accuracy.
Moreover, in this critical region of small $t$, the function $f(t)$ was expanded in a power series
with $f_3$, $f_4$, and $f_{4l}$ coefficients deduced from the known high-$k$ expansion 
by Korobov \cite{Korobov:12} ($\mu'=\mu/m$)
\begin{align}
   f_3 =&\ 8\,\sqrt{\mu'}\,{\cal D}\,, \nonumber \\
   f_{4l} =&\ 16\,\mu'\,{\cal D}\,,  \\
   f_4 =&\ \frac{4}{\mu'^2} \!\! \sum_{(a,x),(b,y)}\!\!
   \bigg \langle \frac{\vec r_{ax}}{r_{ax}^3} \frac{\vec r_{by}}{r_{by}^3}\bigg \rangle_{\!\epsilon}\!\!
   -2\,{\cal D}\,\Big(1 + 4\,\mu' \ln\frac{\mu'}{4}   - 4\,\mu'\Big) \,. \nonumber
\end{align}
The higher order expansion terms are obtained from the fit to numerical values of $f(t)$.
As a test, $f_4=72.114\,86(7)$ calculated using the above formula,
agrees well with $72.0(3)$ obtained from the numerical fit.
In order to perform integration  in Eq.~\eqref{lnk0ft}, we use a polynomial interpolation 
of the integrand for $t > 0.1$, and power expansion for the critical  region $t \in [0, 0.1]$.

The convergence of the Bethe logarithm with the increasing size
of the naECG basis is shown in Table~\ref{Tlnk}. Six significant figures can be considered
stable and the estimated relative accuracy is a half ppm. Our recommended value $\ln k_0=3.018\,304\,9(15)$
is consistent with $3.0188$ obtained in the framework of adiabatic approximation \cite{PLPKPJ09},
and the difference between them divided by $\ln k_0$ is smaller than the electron-proton mass ratio.
An analogous difference of $0.0005$ between adiabatic and nonadiabatic $\ln k_0$ has been noted
for H$_2^+$ \cite{Korobov:06a,PLPKPJ09}.
\begin{table}[t!hb]
\caption{Convergence of the nonrelativistic energy $E$ and the Bethe logarithm $\ln k_0$ 
  with the increasing size $N$ of the naECG basis set. The final uncertainty for $\ln k_0$
  is due to numerical inaccuracy of $f(t)$ at small $t$. }
\label{Tlnk}
\begin{ruledtabular}
\begin{tabular}{cx{3.11}x{4.12}}
\centt{$N$} & \cent{E} & \cent{\ln k_0} \\
\hline\\[-1.5ex]
128               & -1.164\,023\,669\,155   & 3.016\,586\,1 \\
256               & -1.164\,024\,987\,878   & 3.018\,137\,0 \\
512               & -1.164\,025\,027\,334   & 3.018\,258\,91 \\
1024              & -1.164\,025\,030\,593   & 3.018\,301\,73 \\
2048              & -1.164\,025\,030\,843   & 3.018\,303\,90 \\
$\infty$          & -1.164\,025\,030\,86(3) & 3.018\,304\,9(15) \\
\end{tabular}                            
\end{ruledtabular}   
\end{table}

{\sl Higher order QED} --
Because of the significant increase in the accuracy of the QED correction achieved in this work,
the dominating contribution to the uncertainty comes from the higher order $E^{(7)}$ correction.
Currently, an explicit form of this correction is unknown, which prevents its
accurate evaluation. Its first estimation, made within the BO approximation 
framework, was reported in Ref.~\cite{PKCP:16}. Here, we account for several additional terms, namely
\begin{align}
  E^{(7)} \approx&\  \pi\,\Big\langle\sum_{a,x}\delta^3(r_{ax})\Big\rangle
  \bigg\{ \frac{1}{\pi}\big[ A_{60} + A_{61}\,\ln\alpha^{-2} \label{25} \\&\ 
  + A_{62}\, \ln^2\alpha^{-2}\big] + \frac{1}{\pi^2}\,B_{50} + \frac{1}{\pi^3}\,C_{40}\biggr\} - 2\,E^{(7)}({\rm H}),
\nonumber
\end{align}
and assume a conservative 25\% uncertainty.
All the coefficients $A$, $B$, and $C$ can be found in Ref.~\cite{Eides:01}, and we use the values for the $1S$ state of H.
The dominating term is the one containing $A_{62} = -1$ and inclusion of all the other terms
decreases $E^{(7)}$ by about 14\%. 

{\sl Summary} --
Theoretical predictions for all the known contributions to $D_{0,0}$ and $D_{0,1}$ are assembled in Table \ref{TD}.
By $D_{v,J}$ we denote there the dissociation energy of the hydrogen molecule in the state with the vibrational
number $v$ and rotational $J$.
\begin{table*}[!htb]
\renewcommand{\arraystretch}{1.3}
\caption{Theoretical predictions for the dissociation energy budget for
  the ground level of H$_2$. $E^{(6)}_{\rm sec}$ 
  is a second order correction due to relativistic BO potential;
  $E_\mathrm{FS}$ is the finite nuclear size correction with
  $r_p = 0.84087(39)$ fm \cite{Antognini:13}. All the energy entries are given in $\icm$.}
\label{TD}
\begin{ruledtabular}
\begin{tabular}{lx{6.11}x{6.11}x{4.11}l}
\centt{Contribution} & \centt{$D_{0,0}$} & \centt{$D_{0,1}$} & \centt{$(0,1)\rightarrow(0,0)$} & Remarks and references\\
\hline                                                            
$E^{(2)}$ & 36\,118.797\,746\,10(3)& 36\,000.312\,485\,66(2)& 118.485\,260\,44(4)& naJC; \cite{PK:18a} \\
$E^{(4)}$ &      -0.531\,215\,6(5) &      -0.533\,799\,2(5) &   0.002\,583\,56(1)& naECG; \cite{PSKP:18}, this work \\
$E^{(5)}$ &      -0.194\,910\,43(15) &    -0.193\,887\,7(11) &  -0.001\,022\,7(11)& naECG; \cite{PKP:17}, this work  \\
$E^{(6)}$ &      -0.002\,067(6)    &      -0.002\,058(6)    &  -0.000\,008\,9    & BO; \cite{PKCP:16} \\
$E^{(6)}_{\rm sec}$& 0.000\,009\,2  &      0.000\,009\,1    &   0.000\,000\,1    & BO; this work \\
$E^{(7)}$ &       0.000\,101(25)   &       0.000\,101(25)   &   0.000\,000\,5(1)    & BO; \cite{PKCP:16,PKP:17} \\
$E^{(4)}_\mathrm{FS}$&-0.000\,031  &      -0.000\,031       &  -0.000\,000\,2    & BO; \cite{PKCP:16,PKP:17} \\
Total               & 36\,118.069\,632(26)   & 35\,999.582\,820(26)   & 118.486\,812\,7(11)&\\[2ex] 
Exp.	              & 36\,118.069\,62(37)    & 35\,999.582\,894(25)   & 118.486\,8(1)      & \cite{Liu:09}, \cite{Cheng:18}, \cite{Jennings:84} \\
Diff.               &      -0.000\,01(37)    &      0.000\,074(36)    &   0.000\,0(1)      & \\
Exp.	              & 36\,118.069\,45(31)    & 35\,999.582\,834(11)                       &                    & \cite{Altmann:18}, \cite{Holsch:19} \\
Diff.               &      -0.000\,18(31)    &      0.000\,014(28) \\
\end{tabular}                            
\end{ruledtabular}   
\end{table*}
The nonrelativistic contribution was calculated directly for $D_{0,0}$ and $D_{0,1}$.
All the corrections were calculated for $D_{0,0}$ and separately for the rotational 
excitation energy using NAPT. However, the QED correction for this excitation energy
was calculated within the BO approximation only, but the related uncertainty is small. 
Finally, the dissociation energy $D_{0,1}$ was
obtained as the difference between $D_{0,0}$ and the rotational excitation.

The improved theoretical result for the ground state dissociation energy $D_{0,0}$
of the hydrogen molecule is in very good agreement with the most recent measurements 
\cite{Liu:09,Altmann:18}, but their uncertainties are an order of magnitude larger.
The situation is more intriguing for the dissociation energy $D_{0,1}$ of the first 
rotationally excited state. Although, our theoretical prediction differs by 2$\,\sigma$
from the equally accurate recent measurement \cite{Cheng:18}, it is in very good agreement
with the twice as accurate measurement reported in parallel to this work \cite{Holsch:19}, see Table III.

{\sl Conclusions} --
This work concludes efforts to increase the accuracy of theoretical predictions 
for the dissociation energy of H$_2$, feasible within the framework of the existing theory.
Further progress is likely, provided an explicit formula for the $E^{(7)}$ term
in the expansion Eq.~(\ref{alphaexp}) is found.
Currently, the main uncertainty of $25\times 10^{-6}\,\icm$ (0.75 MHz) comes from 
this correction, which for the time being, is estimated using the known atomic hydrogen 
formula, see  Eq. (\ref{25}). Despite this approximation, the new results for dissociation energies
of the hydrogen molecule become the most accurate ever obtained for any molecule.

Regarding the possibility of determination of
the Rydberg constant or the proton charge radius, let us point out that in the atomic hydrogen,
apart from $1S\text{-}2S$ transition, there is no other narrow transition, and the present
charge radius determination relies on an average of many transitions with much larger natural
linewidth than the accuracy of individual  measurements. The alternative route suggested by Merkt \cite{fmerkt}
is to use the ionization energy of the hydrogen molecule,
as a second transition, because its natural width is exactly zero.
The determination of the dissociation energy in Refs. \cite{Liu:09, Cheng:18, Altmann:18, Holsch:19}
is in fact the measurement of the ionization energy $E$\hbox{$({\rm H}_2,{\rm IP})$},
\begin{equation}
E({\rm H}_2,{\rm IP}) = D_0({\rm H}_2) + E({\rm H},{\rm IP}) - D_0({\rm H}^+_2) \label{26} 
\end{equation}  
which for ortho-H$_2$ amounts to about $124\,357$ cm$^{-1}$ \cite{Cheng:18}.
The ratio with the precisely known $2S$-$1S$ transition $82\,259$ cm$^{-1}$  \cite{Parthey:11}
is independent of the Rydberg constant, but depends on the proton charge radius through
\begin{equation}
  \frac{E({\rm H}_2,{\rm IP})}{E({\rm H},2S\text{-}1S)} =  1.512\ldots - 1.4\cdot 10^{-10}\, r_p^2/\mathrm{fm}^2\,.
\end{equation}
Consequently, one needs to achieve
\begin{align} \nonumber
  \delta E({\rm H}_2,{\rm IP}) &= E({\rm H},2S\text{-}1S)\, 2\cdot 0.01\cdot 
  1.4\cdot 10^{-10}\, r_p^2/\mathrm{fm}^2 \\
  &= 1.6\cdot 10^{-7}\,\icm\;(5\, {\rm kHz})
\end{align}    
accuracy for the ionization energy of H$_2$ to obtain the proton radius with 1\% precision.
Among the contributions to $E({\rm H}_2,{\rm IP})$ in Eq. (\ref{26}), the last two
are known with much higher precision \cite{Korobov:17} than required.
Therefore, it is only $D_0({\rm H}_2)$ which needs to be improved.
This can be achieved by the evaluation of the nonadiabatic $E^{(6)}$, $E^{(7)}$ in the BO approximation,
and the $E^{(8)}$ contribution using the atomic hydrogen theory. 
Among them, the calculation of $E^{(7)}$ is the most demanding task,
since it has not yet been accomplished for helium or for any other system except the hydrogen atom
and H$_2^+$ ion \cite{Eides:01,Korobov:14}, but it is feasible using present-day technologies. 

{\sl Acknowledgments} --
This research was supported by National Science Center (Poland) Grant No.  2014/15/B/ST4/05022
as well as by a computing grant from Pozna\'n Supercomputing and Networking Center and by PL-Grid Infrastructure.


\begin{thebibliography}{34}
\expandafter\ifx\csname natexlab\endcsname\relax\def\natexlab#1{#1}\fi
\expandafter\ifx\csname bibnamefont\endcsname\relax
  \def\bibnamefont#1{#1}\fi
\expandafter\ifx\csname bibfnamefont\endcsname\relax
  \def\bibfnamefont#1{#1}\fi
\expandafter\ifx\csname citenamefont\endcsname\relax
  \def\citenamefont#1{#1}\fi
\expandafter\ifx\csname url\endcsname\relax
  \def\url#1{\texttt{#1}}\fi
\expandafter\ifx\csname urlprefix\endcsname\relax\def\urlprefix{URL }\fi
\providecommand{\bibinfo}[2]{#2}
\providecommand{\eprint}[2][]{\url{#2}}

\bibitem[{\citenamefont{Beyer et~al.}(2017)\citenamefont{Beyer, Maisenbacher,
  Matveev, Pohl, Khabarova, Grinin, Lamour, Yost, H{\"a}nsch, Kolachevsky
  et~al.}}]{Beyer:17}
\bibinfo{author}{\bibfnamefont{A.}~\bibnamefont{Beyer}},
  \bibinfo{author}{\bibfnamefont{L.}~\bibnamefont{Maisenbacher}},
  \bibinfo{author}{\bibfnamefont{A.}~\bibnamefont{Matveev}},
  \bibinfo{author}{\bibfnamefont{R.}~\bibnamefont{Pohl}},
  \bibinfo{author}{\bibfnamefont{K.}~\bibnamefont{Khabarova}},
  \bibinfo{author}{\bibfnamefont{A.}~\bibnamefont{Grinin}},
  \bibinfo{author}{\bibfnamefont{T.}~\bibnamefont{Lamour}},
  \bibinfo{author}{\bibfnamefont{D.~C.} \bibnamefont{Yost}},
  \bibinfo{author}{\bibfnamefont{T.~W.} \bibnamefont{H{\"a}nsch}},
  \bibinfo{author}{\bibfnamefont{N.}~\bibnamefont{Kolachevsky}},
  \bibnamefont{et~al.}, \bibinfo{journal}{Science}
  \textbf{\bibinfo{volume}{358}}, \bibinfo{pages}{79} (\bibinfo{year}{2017}).
  
\bibitem[{\citenamefont{Liu et~al.}(2009)\citenamefont{Liu, Salumbides,
  Hollenstein, Koelemeij, Eikema, Ubachs, and Merkt}}]{Liu:09}
\bibinfo{author}{\bibfnamefont{J.}~\bibnamefont{Liu}},
  \bibinfo{author}{\bibfnamefont{E.~J.} \bibnamefont{Salumbides}},
  \bibinfo{author}{\bibfnamefont{U.}~\bibnamefont{Hollenstein}},
  \bibinfo{author}{\bibfnamefont{J.~C.~J.} \bibnamefont{Koelemeij}},
  \bibinfo{author}{\bibfnamefont{K.~S.~E.} \bibnamefont{Eikema}},
  \bibinfo{author}{\bibfnamefont{W.}~\bibnamefont{Ubachs}}, \bibnamefont{and}
  \bibinfo{author}{\bibfnamefont{F.}~\bibnamefont{Merkt}}, \bibinfo{journal}{J.
  Chem. Phys.} \textbf{\bibinfo{volume}{130}}, \bibinfo{eid}{174306}
  (\bibinfo{year}{2009}).

\bibitem[{\citenamefont{Sprecher et~al.}(2011)\citenamefont{Sprecher, Jungen,
  Ubachs, and Merkt}}]{Sprecher:11}
\bibinfo{author}{\bibfnamefont{D.}~\bibnamefont{Sprecher}},
  \bibinfo{author}{\bibfnamefont{C.}~\bibnamefont{Jungen}},
  \bibinfo{author}{\bibfnamefont{W.}~\bibnamefont{Ubachs}}, \bibnamefont{and}
  \bibinfo{author}{\bibfnamefont{F.}~\bibnamefont{Merkt}},
  \bibinfo{journal}{Faraday Discuss.} \textbf{\bibinfo{volume}{150}},
  \bibinfo{pages}{51} (\bibinfo{year}{2011}).

\bibitem[{\citenamefont{Cheng et~al.}(2012)\citenamefont{Cheng, Sun, Pan, Wang,
  Liu, Campargue, and Hu}}]{Cheng:12}
\bibinfo{author}{\bibfnamefont{C.-F.} \bibnamefont{Cheng}},
  \bibinfo{author}{\bibfnamefont{Y.~R.} \bibnamefont{Sun}},
  \bibinfo{author}{\bibfnamefont{H.}~\bibnamefont{Pan}},
  \bibinfo{author}{\bibfnamefont{J.}~\bibnamefont{Wang}},
  \bibinfo{author}{\bibfnamefont{A.-W.} \bibnamefont{Liu}},
  \bibinfo{author}{\bibfnamefont{A.}~\bibnamefont{Campargue}},
  \bibnamefont{and} \bibinfo{author}{\bibfnamefont{S.-M.} \bibnamefont{Hu}},
  \bibinfo{journal}{Phys. Rev. A} \textbf{\bibinfo{volume}{85}},
  \bibinfo{pages}{024501} (\bibinfo{year}{2012}).

\bibitem[{\citenamefont{Niu et~al.}(2014)\citenamefont{Niu, Salumbides,
  Dickenson, Eikema, and Ubachs}}]{Niu:14}
\bibinfo{author}{\bibfnamefont{M.}~\bibnamefont{Niu}},
  \bibinfo{author}{\bibfnamefont{E.}~\bibnamefont{Salumbides}},
  \bibinfo{author}{\bibfnamefont{G.}~\bibnamefont{Dickenson}},
  \bibinfo{author}{\bibfnamefont{K.}~\bibnamefont{Eikema}}, \bibnamefont{and}
  \bibinfo{author}{\bibfnamefont{W.}~\bibnamefont{Ubachs}},
  \bibinfo{journal}{J. Mol. Spectrosc.} \textbf{\bibinfo{volume}{300}},
  \bibinfo{pages}{44 } (\bibinfo{year}{2014}).

\bibitem[{\citenamefont{Cheng et~al.}(2018)\citenamefont{Cheng, Hussels, Niu,
  Bethlem, Eikema, Salumbides, Ubachs, Beyer, H{\"o}lsch, Agner
  et~al.}}]{Cheng:18}
\bibinfo{author}{\bibfnamefont{C.}~\bibnamefont{Cheng}},
  \bibinfo{author}{\bibfnamefont{J.}~\bibnamefont{Hussels}},
  \bibinfo{author}{\bibfnamefont{M.}~\bibnamefont{Niu}},
  \bibinfo{author}{\bibfnamefont{H.~L.} \bibnamefont{Bethlem}},
  \bibinfo{author}{\bibfnamefont{K.~S.~E.} \bibnamefont{Eikema}},
  \bibinfo{author}{\bibfnamefont{E.~J.} \bibnamefont{Salumbides}},
  \bibinfo{author}{\bibfnamefont{W.}~\bibnamefont{Ubachs}},
  \bibinfo{author}{\bibfnamefont{M.}~\bibnamefont{Beyer}},
  \bibinfo{author}{\bibfnamefont{N.~J.} \bibnamefont{H{\"o}lsch}},
  \bibinfo{author}{\bibfnamefont{J.~A.} \bibnamefont{Agner}},
  \bibnamefont{et~al.}, \bibinfo{journal}{Phys. Rev. Lett.}
  \textbf{\bibinfo{volume}{121}}, \bibinfo{pages}{013001}
  (\bibinfo{year}{2018}).

\bibitem[{\citenamefont{Altmann et~al.}(2018)\citenamefont{Altmann, Dreissen,
  Salumbides, Ubachs, and Eikema}}]{Altmann:18}
\bibinfo{author}{\bibfnamefont{R.~K.} \bibnamefont{Altmann}},
  \bibinfo{author}{\bibfnamefont{L.~S.} \bibnamefont{Dreissen}},
  \bibinfo{author}{\bibfnamefont{E.~J.} \bibnamefont{Salumbides}},
  \bibinfo{author}{\bibfnamefont{W.}~\bibnamefont{Ubachs}}, \bibnamefont{and}
  \bibinfo{author}{\bibfnamefont{K.~S.~E.} \bibnamefont{Eikema}},
  \bibinfo{journal}{Phys. Rev. Lett.} \textbf{\bibinfo{volume}{120}},
  \bibinfo{pages}{043204} (\bibinfo{year}{2018}).

\bibitem[{\citenamefont{Pachucki and Komasa}(2009)}]{PK09}
\bibinfo{author}{\bibfnamefont{K.}~\bibnamefont{Pachucki}} \bibnamefont{and}
  \bibinfo{author}{\bibfnamefont{J.}~\bibnamefont{Komasa}},
  \bibinfo{journal}{J. Chem. Phys.} \textbf{\bibinfo{volume}{130}},
  \bibinfo{pages}{164113} (\bibinfo{year}{2009}).

\bibitem[{\citenamefont{Pachucki and Komasa}(2015)}]{PK15}
\bibinfo{author}{\bibfnamefont{K.}~\bibnamefont{Pachucki}} \bibnamefont{and}
  \bibinfo{author}{\bibfnamefont{J.}~\bibnamefont{Komasa}},
  \bibinfo{journal}{J. Chem. Phys.} \textbf{\bibinfo{volume}{143}},
  \bibinfo{eid}{034111} (\bibinfo{year}{2015}).

\bibitem[{\citenamefont{Pachucki and Komasa}(2016)}]{PK:16}
\bibinfo{author}{\bibfnamefont{K.}~\bibnamefont{Pachucki}} \bibnamefont{and}
  \bibinfo{author}{\bibfnamefont{J.}~\bibnamefont{Komasa}},
  \bibinfo{journal}{J. Chem. Phys.} \textbf{\bibinfo{volume}{144}},
  \bibinfo{eid}{164306} (\bibinfo{year}{2016}).

\bibitem[{\citenamefont{Wang and Yan}(2018{\natexlab{a}})}]{Wang:18a}
\bibinfo{author}{\bibfnamefont{L.~M.} \bibnamefont{Wang}} \bibnamefont{and}
  \bibinfo{author}{\bibfnamefont{Z.-C.} \bibnamefont{Yan}},
  \bibinfo{journal}{Phys. Rev. A} \textbf{\bibinfo{volume}{97}},
  \bibinfo{pages}{060501} (\bibinfo{year}{2018}{\natexlab{a}}).

\bibitem[{\citenamefont{Puchalski et~al.}(2018)\citenamefont{Puchalski,
  Spyszkiewicz, Komasa, and Pachucki}}]{PSKP:18}
\bibinfo{author}{\bibfnamefont{M.}~\bibnamefont{Puchalski}},
  \bibinfo{author}{\bibfnamefont{A.}~\bibnamefont{Spyszkiewicz}},
  \bibinfo{author}{\bibfnamefont{J.}~\bibnamefont{Komasa}}, \bibnamefont{and}
  \bibinfo{author}{\bibfnamefont{K.}~\bibnamefont{Pachucki}},
  \bibinfo{journal}{Phys. Rev. Lett.} \textbf{\bibinfo{volume}{121}},
  \bibinfo{pages}{073001} (\bibinfo{year}{2018}).

\bibitem[{\citenamefont{Wang and Yan}(2018{\natexlab{b}})}]{Wang:18b}
\bibinfo{author}{\bibfnamefont{L.}~\bibnamefont{Wang}} \bibnamefont{and}
  \bibinfo{author}{\bibfnamefont{Z.-C.} \bibnamefont{Yan}},
  \bibinfo{journal}{Phys. Chem. Chem. Phys.} \textbf{\bibinfo{volume}{20}},
  \bibinfo{pages}{23948} (\bibinfo{year}{2018}{\natexlab{b}}).

\bibitem[{\citenamefont{Puchalski et~al.}(2017)\citenamefont{Puchalski, Komasa,
  and Pachucki}}]{PKP:17}
\bibinfo{author}{\bibfnamefont{M.}~\bibnamefont{Puchalski}},
  \bibinfo{author}{\bibfnamefont{J.}~\bibnamefont{Komasa}}, \bibnamefont{and}
  \bibinfo{author}{\bibfnamefont{K.}~\bibnamefont{Pachucki}},
  \bibinfo{journal}{Phys. Rev. A} \textbf{\bibinfo{volume}{95}},
  \bibinfo{pages}{052506} (\bibinfo{year}{2017}).

\bibitem[{\citenamefont{Czachorowski et~al.}(2018)\citenamefont{Czachorowski,
  Puchalski, Komasa, and Pachucki}}]{CPKP:18}
\bibinfo{author}{\bibfnamefont{P.}~\bibnamefont{Czachorowski}},
  \bibinfo{author}{\bibfnamefont{M.}~\bibnamefont{Puchalski}},
  \bibinfo{author}{\bibfnamefont{J.}~\bibnamefont{Komasa}}, \bibnamefont{and}
  \bibinfo{author}{\bibfnamefont{K.}~\bibnamefont{Pachucki}},
  \bibinfo{journal}{Phys. Rev. A} \textbf{\bibinfo{volume}{98}},
  \bibinfo{pages}{052506} (\bibinfo{year}{2018}).

\bibitem[{\citenamefont{Piszczatowski et~al.}(2009)\citenamefont{Piszczatowski,
  Lach, Przybytek, Komasa, Pachucki, and Jeziorski}}]{PLPKPJ09}
\bibinfo{author}{\bibfnamefont{K.}~\bibnamefont{Piszczatowski}},
  \bibinfo{author}{\bibfnamefont{G.}~\bibnamefont{Lach}},
  \bibinfo{author}{\bibfnamefont{M.}~\bibnamefont{Przybytek}},
  \bibinfo{author}{\bibfnamefont{J.}~\bibnamefont{Komasa}},
  \bibinfo{author}{\bibfnamefont{K.}~\bibnamefont{Pachucki}}, \bibnamefont{and}
  \bibinfo{author}{\bibfnamefont{B.}~\bibnamefont{Jeziorski}},
  \bibinfo{journal}{J. Chem. Theory Comput.} \textbf{\bibinfo{volume}{5}},
  \bibinfo{pages}{3039} (\bibinfo{year}{2009}).

\bibitem[{\citenamefont{Puchalski et~al.}(2016)\citenamefont{Puchalski, Komasa,
  Czachorowski, and Pachucki}}]{PKCP:16}
\bibinfo{author}{\bibfnamefont{M.}~\bibnamefont{Puchalski}},
  \bibinfo{author}{\bibfnamefont{J.}~\bibnamefont{Komasa}},
  \bibinfo{author}{\bibfnamefont{P.}~\bibnamefont{Czachorowski}},
  \bibnamefont{and} \bibinfo{author}{\bibfnamefont{K.}~\bibnamefont{Pachucki}},
  \bibinfo{journal}{Phys. Rev. Lett.} \textbf{\bibinfo{volume}{117}},
  \bibinfo{pages}{263002} (\bibinfo{year}{2016}).

\bibitem[{\citenamefont{Eides et~al.}(2001)\citenamefont{Eides, Grotch, and
  Shelyuto}}]{Eides:01}
\bibinfo{author}{\bibfnamefont{M.~I.} \bibnamefont{Eides}},
  \bibinfo{author}{\bibfnamefont{H.}~\bibnamefont{Grotch}}, \bibnamefont{and}
  \bibinfo{author}{\bibfnamefont{V.~A.} \bibnamefont{Shelyuto}},
  \bibinfo{journal}{Phys. Rep.} \textbf{\bibinfo{volume}{342}},
  \bibinfo{pages}{63 } (\bibinfo{year}{2001}).

\bibitem[{\citenamefont{Pachucki and Sapirstein}(2000)}]{Pachucki:00}
\bibinfo{author}{\bibfnamefont{K.}~\bibnamefont{Pachucki}} \bibnamefont{and}
  \bibinfo{author}{\bibfnamefont{J.}~\bibnamefont{Sapirstein}},
  \bibinfo{journal}{J. Phys. B} \textbf{\bibinfo{volume}{33}},
  \bibinfo{pages}{455} (\bibinfo{year}{2000}).

\bibitem[{\citenamefont{Drake and Swainson}(1990)}]{Drake:90}
\bibinfo{author}{\bibfnamefont{G.~W.~F.} \bibnamefont{Drake}} \bibnamefont{and}
  \bibinfo{author}{\bibfnamefont{R.~A.} \bibnamefont{Swainson}},
  \bibinfo{journal}{Phys. Rev. A} \textbf{\bibinfo{volume}{41}},
  \bibinfo{pages}{1243} (\bibinfo{year}{1990}).

\bibitem[{\citenamefont{Wiringa et~al.}(1995)\citenamefont{Wiringa, Stoks, and
  Schiavilla}}]{Wiringa:95}
\bibinfo{author}{\bibfnamefont{R.~B.} \bibnamefont{Wiringa}},
  \bibinfo{author}{\bibfnamefont{V.~G.~J.} \bibnamefont{Stoks}},
  \bibnamefont{and}
  \bibinfo{author}{\bibfnamefont{R.}~\bibnamefont{Schiavilla}},
  \bibinfo{journal}{Phys. Rev. C} \textbf{\bibinfo{volume}{51}},
  \bibinfo{pages}{38} (\bibinfo{year}{1995}).

\bibitem[{\citenamefont{Pohl et~al.}(2010)\citenamefont{Pohl, Antognini, Nez,
  Amaro, Biraben, Cardoso, Covita, Dax, Dhawan, Fernandes et~al.}}]{pohl:10}
\bibinfo{author}{\bibfnamefont{R.}~\bibnamefont{Pohl}},
  \bibinfo{author}{\bibfnamefont{A.}~\bibnamefont{Antognini}},
  \bibinfo{author}{\bibfnamefont{F.}~\bibnamefont{Nez}},
  \bibinfo{author}{\bibfnamefont{F.~D.} \bibnamefont{Amaro}},
  \bibinfo{author}{\bibfnamefont{F.}~\bibnamefont{Biraben}},
  \bibinfo{author}{\bibfnamefont{J.~M.~R.} \bibnamefont{Cardoso}},
  \bibinfo{author}{\bibfnamefont{D.~S.} \bibnamefont{Covita}},
  \bibinfo{author}{\bibfnamefont{A.}~\bibnamefont{Dax}},
  \bibinfo{author}{\bibfnamefont{S.}~\bibnamefont{Dhawan}},
  \bibinfo{author}{\bibfnamefont{L.~M.~P.} \bibnamefont{Fernandes}},
  \bibnamefont{et~al.}, \bibinfo{journal}{Nature}
  \textbf{\bibinfo{volume}{466}}, \bibinfo{pages}{213} (\bibinfo{year}{2010}).

\bibitem[{\citenamefont{Pachucki}(1999)}]{Pachucki:99}
\bibinfo{author}{\bibfnamefont{K.}~\bibnamefont{Pachucki}},
  \bibinfo{journal}{Phys. Rev. A} \textbf{\bibinfo{volume}{60}},
  \bibinfo{pages}{3593} (\bibinfo{year}{1999}).

\bibitem[{\citenamefont{Pachucki et~al.}(2005)\citenamefont{Pachucki, Cencek,
  and Komasa}}]{PCK:05}
\bibinfo{author}{\bibfnamefont{K.}~\bibnamefont{Pachucki}},
  \bibinfo{author}{\bibfnamefont{W.}~\bibnamefont{Cencek}}, \bibnamefont{and}
  \bibinfo{author}{\bibfnamefont{J.}~\bibnamefont{Komasa}},
  \bibinfo{journal}{J. Chem. Phys.} \textbf{\bibinfo{volume}{122}},
  \bibinfo{eid}{184101} (\bibinfo{year}{2005}).

\bibitem[{\citenamefont{Zhou et~al.}(2006)\citenamefont{Zhou, Zhu, and
  Yan}}]{Zhou:06}
\bibinfo{author}{\bibfnamefont{B.-L.} \bibnamefont{Zhou}},
  \bibinfo{author}{\bibfnamefont{J.-M.} \bibnamefont{Zhu}}, \bibnamefont{and}
  \bibinfo{author}{\bibfnamefont{Z.-C.} \bibnamefont{Yan}},
  \bibinfo{journal}{Phys. Rev. A} \textbf{\bibinfo{volume}{73}},
  \bibinfo{pages}{014501} (\bibinfo{year}{2006}).

\bibitem[{\citenamefont{Korobov}(2012)}]{Korobov:12}
\bibinfo{author}{\bibfnamefont{V.~I.} \bibnamefont{Korobov}},
  \bibinfo{journal}{Phys. Rev. A} \textbf{\bibinfo{volume}{85}},
  \bibinfo{pages}{042514} (\bibinfo{year}{2012}).

\bibitem[{\citenamefont{Korobov}(2006)}]{Korobov:06a}
\bibinfo{author}{\bibfnamefont{V.~I.} \bibnamefont{Korobov}},
  \bibinfo{journal}{Phys. Rev. A} \textbf{\bibinfo{volume}{73}},
  \bibinfo{pages}{024502} (\bibinfo{year}{2006}).

\bibitem[{\citenamefont{Antognini et~al.}(2013)\citenamefont{Antognini, Nez,
  Schuhmann, Amaro, Biraben, Cardoso, Covita, Dax, Dhawan, Diepold
  et~al.}}]{Antognini:13}
\bibinfo{author}{\bibfnamefont{A.}~\bibnamefont{Antognini}},
  \bibinfo{author}{\bibfnamefont{F.}~\bibnamefont{Nez}},
  \bibinfo{author}{\bibfnamefont{K.}~\bibnamefont{Schuhmann}},
  \bibinfo{author}{\bibfnamefont{F.~D.} \bibnamefont{Amaro}},
  \bibinfo{author}{\bibfnamefont{F.}~\bibnamefont{Biraben}},
  \bibinfo{author}{\bibfnamefont{J.~M.~R.} \bibnamefont{Cardoso}},
  \bibinfo{author}{\bibfnamefont{D.~S.} \bibnamefont{Covita}},
  \bibinfo{author}{\bibfnamefont{A.}~\bibnamefont{Dax}},
  \bibinfo{author}{\bibfnamefont{S.}~\bibnamefont{Dhawan}},
  \bibinfo{author}{\bibfnamefont{M.}~\bibnamefont{Diepold}},
  \bibnamefont{et~al.}, \bibinfo{journal}{Science}
  \textbf{\bibinfo{volume}{339}}, \bibinfo{pages}{417} (\bibinfo{year}{2013}).

\bibitem[{\citenamefont{Pachucki and Komasa}(2018)}]{PK:18a}
\bibinfo{author}{\bibfnamefont{K.}~\bibnamefont{Pachucki}} \bibnamefont{and}
  \bibinfo{author}{\bibfnamefont{J.}~\bibnamefont{Komasa}},
  \bibinfo{journal}{Phys. Chem. Chem. Phys.} \textbf{\bibinfo{volume}{20}},
  \bibinfo{pages}{247} (\bibinfo{year}{2018}).
  
\bibitem[{\citenamefont{{Jennings} et~al.}(1984)\citenamefont{{Jennings},
  {Bragg}, and {Brault}}}]{Jennings:84}
\bibinfo{author}{\bibfnamefont{D.~E.} \bibnamefont{{Jennings}}},
  \bibinfo{author}{\bibfnamefont{S.~L.} \bibnamefont{{Bragg}}},
  \bibnamefont{and} \bibinfo{author}{\bibfnamefont{J.~W.}
  \bibnamefont{{Brault}}}, \bibinfo{journal}{Astrophys. J.}
  \textbf{\bibinfo{volume}{282}}, \bibinfo{pages}{L85} (\bibinfo{year}{1984}).

\bibitem{fmerkt} F. Merkt, {\em private communication}.
  
\bibitem{Holsch:19}
  N. H\"olsch, M. Beyer, E. Salumbides, K.S.E. Eikema, W.~Ubachs, Ch. Jungen
  and F. Merkt, Phys. Rev. Lett. {\bf 122}, 103002 (2019).

\bibitem[{\citenamefont{Parthey et~al.}(2011)\citenamefont{Parthey, Matveev,
  Alnis, Bernhardt, Beyer, Holzwarth, Maistrou, Pohl, Predehl, Udem
  et~al.}}]{Parthey:11}
\bibinfo{author}{\bibfnamefont{C.~G.} \bibnamefont{Parthey}},
  \bibinfo{author}{\bibfnamefont{A.}~\bibnamefont{Matveev}},
  \bibinfo{author}{\bibfnamefont{J.}~\bibnamefont{Alnis}},
  \bibinfo{author}{\bibfnamefont{B.}~\bibnamefont{Bernhardt}},
  \bibinfo{author}{\bibfnamefont{A.}~\bibnamefont{Beyer}},
  \bibinfo{author}{\bibfnamefont{R.}~\bibnamefont{Holzwarth}},
  \bibinfo{author}{\bibfnamefont{A.}~\bibnamefont{Maistrou}},
  \bibinfo{author}{\bibfnamefont{R.}~\bibnamefont{Pohl}},
  \bibinfo{author}{\bibfnamefont{K.}~\bibnamefont{Predehl}},
  \bibinfo{author}{\bibfnamefont{T.}~\bibnamefont{Udem}}, \bibnamefont{et~al.},
  \bibinfo{journal}{Phys. Rev. Lett.} \textbf{\bibinfo{volume}{107}},
  \bibinfo{pages}{203001} (\bibinfo{year}{2011}).
  
\bibitem[{\citenamefont{Korobov et~al.}(2017)\citenamefont{Korobov, Hilico, and
  Karr}}]{Korobov:17}
\bibinfo{author}{\bibfnamefont{V.~I.} \bibnamefont{Korobov}},
  \bibinfo{author}{\bibfnamefont{L.}~\bibnamefont{Hilico}}, \bibnamefont{and}
  \bibinfo{author}{\bibfnamefont{J.-P.} \bibnamefont{Karr}},
  \bibinfo{journal}{Phys. Rev. Lett.} \textbf{\bibinfo{volume}{118}},
  \bibinfo{pages}{233001} (\bibinfo{year}{2017}).

\bibitem[{\citenamefont{Korobov et~al.}(2014)\citenamefont{Korobov, Hilico, and
  Karr}}]{Korobov:14}
\bibinfo{author}{\bibfnamefont{V.~I.} \bibnamefont{Korobov}},
  \bibinfo{author}{\bibfnamefont{L.}~\bibnamefont{Hilico}}, \bibnamefont{and}
  \bibinfo{author}{\bibfnamefont{J.-P.} \bibnamefont{Karr}},
  \bibinfo{journal}{Phys. Rev. Lett.} \textbf{\bibinfo{volume}{112}},
  \bibinfo{pages}{103003} (\bibinfo{year}{2014}).

\end{thebibliography}

\end{document}